\documentclass{article}
\usepackage{ijcai18}

\usepackage{times}
\usepackage{xcolor}
\usepackage{soul}
\usepackage[utf8]{inputenc}
\usepackage[small]{caption}
\usepackage{graphicx}

\title{Evaluation as a Service architecture and crowdsourced problems solving implemented in Optil.io platform}

\author{
Szymon Wasik$^{1,2}$, 
Maciej Antczak$^1$,
Jan Badura$^1$,
Artur Laskowski$^1$, 
\\ 
$^1$ Institute of Computing Science, Poznan University of Technology, Poland \\
$^2$ Institute of Bioorganic Chemistry, Polish Academy of Sciences, Poland \\
szymon.wasik@cs.put.poznan.pl,
maciej.antczak@cs.put.poznan.pl,\\
jan.badura@cs.put.poznan.pl,
artur.laskowski@cs.put.poznan.pl
}
\begin{document}

\maketitle

\begin{abstract}
Reliable and trustworthy evaluation of algorithms is a challenging process. Firstly, each algorithm has its strengths and weaknesses, and the selection of test instances can significantly influence the assessment process. Secondly, the measured performance of the algorithm highly depends on the test environment architecture, i.e., CPU model, available memory, cache configuration, operating system's kernel, and even compilation flags. Finally, it is often difficult to compare algorithm with software prepared by other researchers. Evaluation as a Service (EaaS) is a cloud computing architecture that tries to make assessment process more reliable by providing online tools and test instances dedicated to the evaluation of algorithms. One of such platforms is Optil.io which gives the possibility to define problems, store evaluation data and evaluate solutions submitted by researchers in almost real time. In this paper, we briefly present this platform together with four challenges that were organized with its support.
\end{abstract}

\section{Introduction} 

Online judges are systems that provide fully automated evaluation of algorithms, which solve computational problems submitted by their users. The term online judge was introduced by Kurnia, Lim, and Cheang in 2001 as an online platform that supports fully automated, real-time evaluation of source code, binaries, or even textual output submitted by participants competing in the particular challenge \cite{Kurnia_2001}. Actually, the development of online judge systems boasts a much longer history, dating back to 1961 when they emerged at Stanford University \cite{Forsythe_1965}. In the past, their most popular application was supporting organization of competitive programming contests such as ACM International Collegiate Programming Contest (ICPC) and International Olympiad in Informatics (IOI) and archiving of problems used during such competitions \cite{Khera_1993}. However, nowadays, their application is much wider, including organization of challenges dedicated to solving data science and optimization problems.

Formally, the online judge system is an online service performing any of the following steps of the evaluation procedure in the cloud computing infrastructure \cite{Wasik_2018}:
\begin{enumerate} 
  \item collects, compiles sources if needed, and verifies the executability of the resultant binary $b$;
  \item assesses solution $b$ based on a set of specific test cases, $T$, defined for the particular computational problem $\Pi$ in a reliable, homogeneous, evaluation environment;
  \item computes the aggregated status $s$ and the evaluation score $v$ based on the statuses and scores of all considered test cases. 
\end{enumerate}

The most popular application of online judge systems is archiving problems from algorithmic contests. The first system in this category that gained significant popularity was UVa Online Judge, followed by many others such as Codeforces, SPOJ, TopCoder, POJ, etc. \cite{Wasik_2018}. The number of programming problems stored by popular online judge systems is so large that even some methods were created to classify them manually (e.g., uHunt) or automatically \cite{Yoon_2006}. Another popular application of online judge systems is the support of education purposes, e.g., in computing science in teaching programming or algorithms and data structures, such as CheckIO, CodinGame or Codeboard. There are even online judges that extend e-learning platforms such as Moodle. Recruiters’ work can also be supported by the online judges that are usually used to verify the programming skills of candidates for employers such as HackerRank or Qualified. Finally, there are development platforms such as DOMjudge or Mooshak that allow integrating selected mechanism provided by the online judge system into custom web service. Detailed classification and features provided by various online judges are described in \cite{Ihantola2015,Combefis_2014,Wasik_2018}.

The approach offered by the online judge platforms has such a large impact that the concept they utilize has already been named as a cloud-based evaluation, or, strictly following the cloud computing naming scheme, as Evaluation as a Service (EaaS). At least two international workshops devoted to this topic have been organized recently \cite{Muller_2016,Hopfgartner_2015}.

\section{Methods - Optil.io platform}

In this paper we would like to shortly present how we used the Evaluation as a Service architecture to implement the Optil.io platform that is focused on evaluating data science and optimization problems \cite{Wasik_2016}. The utilization of EaaS methodology implemented in the Optil.io platform from the user perspective is following:
\begin{enumerate}
  \item User submits the algorithm solving a particular computational problem through a web interface. The algorithm can be submitted in the form of a source code that will be compiled in the provided computational infrastructure or a binary executable.
  \item The platform verifies if the submission can be properly executed and in the case of a source code it verifies if it compiles successfully.
  \item The submission is executed using a homogeneous, safe computing infrastructure on the benchmark including dedicated test cases. During the execution, the evaluation engine verifies if the submission does not exceed strict resource limitations (maximal processing time, RAM utilization, disk storage limit) and does not induce runtime errors. 
  \item Based on the results calculated during the evaluation the ranking of all considered submissions is presented.
\end{enumerate}

An important feature of the EaaS methodology is the possibility to reliably as well as continuously evaluate submissions, at any moment of time, whenever they are submitted. This aspect of EaaS differs from the simple web form that allows only for collecting of submissions through the Internet. The EaaS approach that has been developed as a part of the Optil.io platform provides a reliable and continuous evaluation of algorithms solving complex optimization and data science problems. 

As web platforms implemented according to EaaS architecture can be easily used by users to share their solutions for various problems, including those originating from data science and optimization area, they are often extended to utilize crowdsourcing concept. The term \textit{crowdsourcing} was introduced in 2006 by Jeff Howe \cite{Howe_2006}. However, the concept of crowdsourcing understood as outsourcing work to a vast, usually unnamed, network of people in the form of an open call is quite old. One of its first applications was the discovery of a method for measuring the longitude of a ship in 1714, for which a prize was offered by the British government. Since that time, the concept of crowdsourcing has been utilized many times, but its rapid progression has started after the development of the Internet in the 1990s \cite{Wasik_2015}. The successful examples of its applications are services such as Wikipedia or OpenStreetMap. An extended review of crowdsourcing systems available on the world wide web (WWW) can be found in the survey by Doan et al. \cite{Doan_2011}, and a discussion about the nature of crowdsourcing can be found in the paper by Estelles and Gonzalez \cite{Estelles-Arolas_2012}. 

Crowdsourcing can bring many benefits. As observed by Francis Galton in 1907, the collective opinion of a crowd of individuals can be much more precise than the opinion of any single individual from the crowd. It is a basic assumption of the so-called "wisdom-of-the-crowd" \cite{Galton_1907}, which later evolved into an idea of collective intelligence (CI). A CI is a more general concept defined usually as an intelligence emerging from the collaboration, collective efforts, and competition of many individuals. In recent years, many platforms supporting the crowdsourcing have been implemented, such as InnoCentive or CrowdAnalytix.

On the other hand, programming challenges can be very successful in solving complex science- and industry-inspired computing problems. This has been proven especially in the data mining field by the Kaggle platform \cite{Dhar_2013}. However, there are also many other successful utilizations of such events, e.g., Dream Challenges, ROADEF Challenge, VeRoLog Solver Challenge or TopCoder. A typical challenge organization requires, first, publishing a description of the challenge on the Internet and next, collecting submissions from the crowd of practitioners. Submissions are usually up-loaded as binaries executed by judges after the challenge submission deadline or textual output files generated by contestants executing their own code on a predefined benchmark set of test cases provided by the judges. A very similar approach was implemented in Optil.io platform which make possible not only to submit the algorithms and evaluate them in the cloud, but also to organize programming challenges which objective is to solve difficult optimization or data science problems.

Challenges organized at Optil.io platform employ the continuous evaluation built on EaaS architecture. Participants can submit any source code developed in any programming language. To prevent overfitting of solutions to test instances, we can divide tests into public and private sets. To ensure smooth communication between algorithms and the evaluation engine we developed a simple communication channel based on the standard input/output. Privacy and intellectual property rights to the data submitted by users are regulated by terms of service, challenges rules, and privacy policy prepared by the cooperating lawyer and accepted by all participants of the challenge. In general, they leave the intellectual property to the author and only the winner has to provide his solution to the challenge organizer in order to receive the prize.

\section{Results}

\begin{figure*}[ht!]
  \centering
  \includegraphics[width=0.75\textwidth]{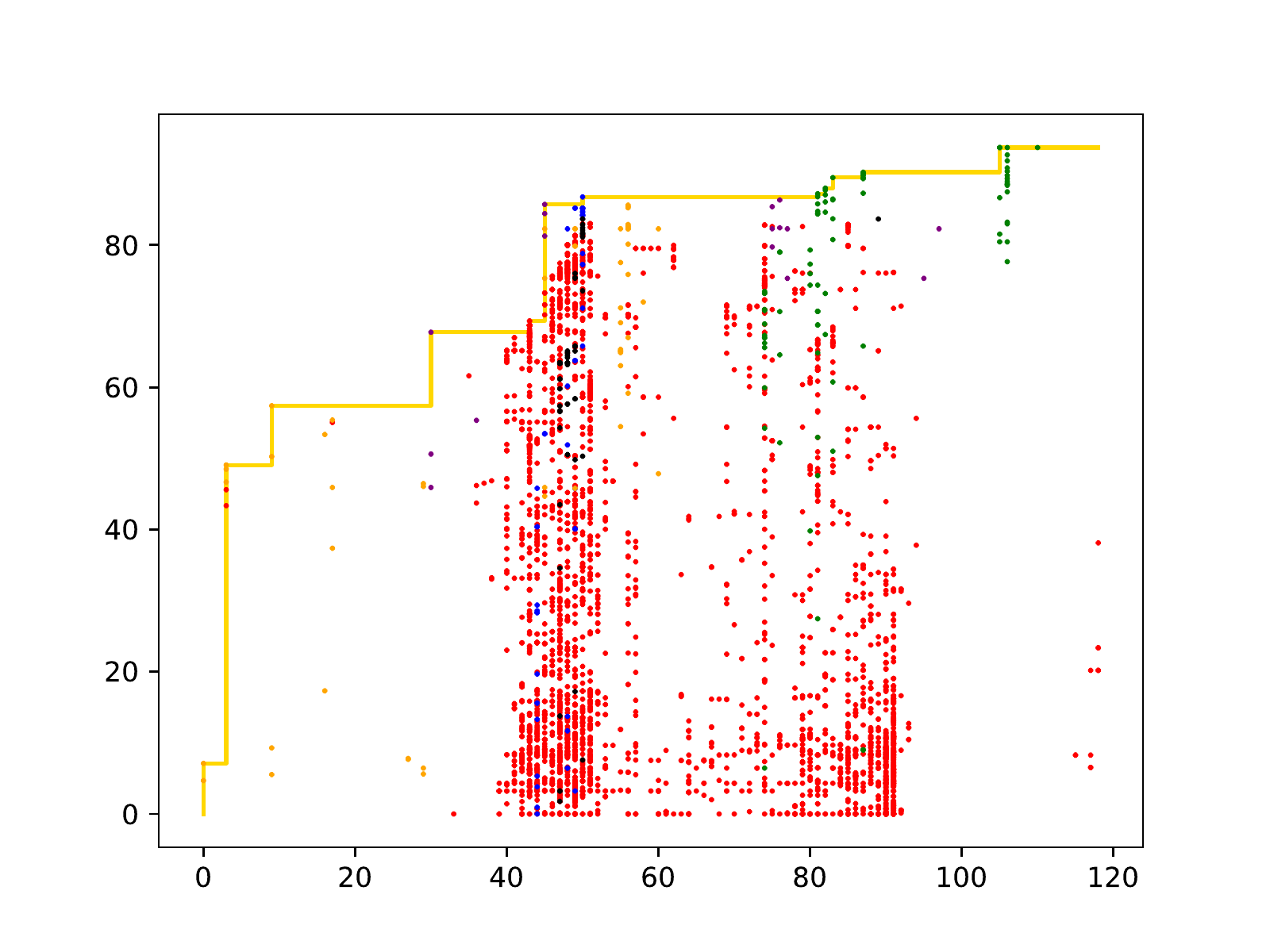}
  \caption{The results of the challenge dedicated to solving the variant of an orienteering problem. The horizontal axis presents the progress of the challenge in days. The vertical axis presents the total, aggregated score obtained by the solutions (relatively to the best submission which is assigned 100 points). Each data point represents the single submission. The yellow line presents how the score obtained by the best solution increased in time. Top 5 ranked in the final standing contestants are marked with colors different than red.\label{fig:ranking}}
\end{figure*}

We have already organized four challenges using Optil.io platform. Two of them were focused on simple combinatorial problems - variants of facility location and orienteering problems. The remaining two challenges were dedicated to supporting PACE Challenge organized in collaboration with other European universities: Saarland University, Friedrich-Schiller-University Jena, and University Paris Dauphine.

The goal of the Parameterized Algorithms and Computational Experiments (PACE) Challenge is to investigate the applicability of algorithmic ideas studied and developed in the subfields of multivariate, fine-grained, parameterized, or fixed-parameter tractable algorithms. The 2017 and 2018 editions of PACE were hosted using Optil.io platform. Up to this point, the organizers run solutions submitted by users only once, at the end of the contest, using manually executed scripts. Optil.io platform provided them with automatic, continuous evaluation method. Contestants could submit their algorithms to the platform and have them assessed almost instantly using 200 instances. 100 of them were public ones, for which participants could see their results in real time and compare those results with others. Another 100 instances were private. Results of evaluation using these instances were visible only to organizers, and after the challenge deadline, they were used to determine the winner. For each instance algorithm could use up to 30 minutes of processor time, thus requiring to perform 100 hours of computation per submission. Thanks to the parallelization, contestants could check their results on public instances after just around 90 minutes. 

Organization of all of the challenges allowed observing many interesting behaviors emerging from the crowdsourcing approach. We could observe how contestants worked on their solutions improving them, how much time they required to find the optimal solution for particular instances and how the best results obtained by any of the contestants changed in time. An example plot presenting the progress of the challenge is presented in Figure \ref{fig:ranking}.

\section{Acknowledgments}

All authors were supported by the National Center for Research and Development, Poland [grant no. LIDER/004/103/L-5/13/NCBR/2014]. Moreover, development tools used during the Optil.io project (JIRA and Bitbucket) were shared by PLGrid infrastrucutre. 

\bibliographystyle{named}
\bibliography{ijcai18}

\end{document}